\documentclass[namedreferences]{solarphysics}
\usepackage[optionalrh,solaromanenum,natbib]{spr-sola-addons} 
\usepackage{graphicx}                    
\usepackage{courier}                    	
\usepackage{amssymb}                   
\usepackage{color}                       	
\usepackage{url}                         	

\newcommand{\Ha}{H$\alpha$}
\newcommand{\kms}{km~s$^{-1}$}
\newcommand{\dhline}{\hline & \\[-1.7em]\hline}
\newcommand{\NiI}{\hbox{{\rm Ni}\kern 0.1em{\sc i}}}
\newcommand{\FeI}{\hbox{{\rm Fe}\kern 0.1em{\sc i}}}

\begin{document}

\begin{article}

\begin{opening}

\title{The Origin of Sequential Chromospheric Brightenings}

%
\author{M.S.~\surname{Kirk}$^{1,2}$\sep
        K.S.~\surname{Balasubramaniam}$^{3}$\sep
        J.~\surname{Jackiewicz}$^{2}$\sep
        H.R.~\surname{Gilbert}$^{1}$    
       }

%
\runningauthor{M.S. Kirk {\it et al.}}
\runningtitle{The Origin of SCBs}

%
  \institute{$^{1}$ NASA Goddard Space Flight Center, Code 670, Greenbelt, MD 20771, USA\\
		email:~\url{michael.s.kirk@nasa.gov} \\
		$^{2}$ Department of Astronomy, New Mexico State University, Las Cruces, NM 88003, USA\\
              $^{3}$  AFRL Battlespace Environment Division (AFRL/RVB), Albuquerque, NM 87123, USA\\
             }

\begin{abstract}
Sequential chromospheric brightenings (SCBs) are often observed in the immediate vicinity of erupting flares and are associated with coronal mass ejections. Since their initial discovery in 2005, there have been several subsequent investigations of SCBs. These studies have used differing detection and analysis techniques, making it difficult to compare results between studies. This work employs the automated detection algorithm of Kirk et al. (Solar Phys. 283, 97, 2013) to extract the physical characteristics of SCBs in 11 flares of varying size and intensity. We demonstrate that the magnetic substructure within the SCB appears to have a significantly smaller area than the corresponding \Ha\ emission. We conclude that SCBs originate in the lower corona around $0.1~R_\odot$ above the photosphere, propagate away from the flare center at speeds of $35-85$~\kms, and have peak photosphere magnetic intensities of $148\pm2.9$~G. In light of these measurements, we infer SCBs to be distinctive chromospheric signatures of erupting coronal mass ejections. 
\end{abstract}

%

\end{opening}


 \section{Introduction to SCBs}
	\label{s:intro}
Two ribbon chromospheric flares observed in \Ha\ appear well-organized when first examined: ribbons impulsively brighten, separate, and exponentially decay back to pre-flare levels. Upon closer inspection of \Ha\ flares, there is often a significant number of compact brightenings in concert with the flare eruption that are spatially separated from the evolving flare ribbon. One class of these off-ribbon chromospheric brightening was first classified by~\citet{Bala2005} in a 19 December 2002 M2.7 flare. Using a multi-wavelength data set to analyze the 19 December flare and eruption of a large scale transequatorial loop, \citet{Bala2005} observed a large scale coronal dimming, flares in both the north and south hemispheres, and a halo coronal mass ejection (CME). In \Ha\ images of the same event, the loop eruption manifested itself as flare precursor-brightenings, sympathetic flares, and cospatial propagating chromospheric brightenings. Termed sequential chromospheric brightenings (SCBs), the speed of this propagating disturbance was measured to be between $600-800$~\kms. Although the disturbance propagated at similar speeds to extreme ultraviolet (EUV) flare waves, they differed from typical waves observed in \Ha\ (Moreton waves) in that they were not observed in off-band images, had an angular propagation width of less than 30$^{\circ}$, and appeared as distinctly individual points of brightening instead of continuous fronts~\citep{Kirk2012b}. 

\citet{Kirk2011} refined the technique and developed a systematic method for identifying and measuring properties of SCBs by employing an automated detection and tracking algorithm. They concluded that SCBs originate during the impulsive rise phase of the flare and often precede the \Ha\ flare intensity peak. ~\citet{Kirk2012b,Kirk2012a} discovered that the nature of SCBs are phenomenologically distinct from other compact brighteings observed in the chromosphere due to their impulsive intensity signatures, unique Doppler velocity profiles, and origin in the impulsive phase of flare \Ha\ intensity evolution.  \citet{Kirk_PhD} and~\citet{Kirk2014} found that SCBs are a type of localized chromospheric heating and ablation due to impacting coronal plasma and have associated temperatures of $T\approx10^{5}$~K. As an ensemble, SCBs are tracked to propagate outward, away from the flare center, at velocities between $41-89$~\kms~\citep{Kirk2012a}.

Between the initial parametrization of SCBs in 2005 and 2007, and contemporary work completed in the past few years, several inconsistencies have emerged in the stated characteristics of SCBs. Specifically, \citet{Bala2005} found SCB propagation speeds to be between 600--800~\kms\ while \citet{Kirk2012a} found more modest speeds of $41-89$~\kms.  Also, \citet{Bala2006} found SCBs to be related to their host flare only in 65\% of the cases studied and postulated that ``...SCBs are not a direct consequence of flares,'' which differs from the empirical models of~\citet{Kirk2012a} and~\citet{Pevtsov2007}. Furthermore, both \citet{Bala2006} and \citet{Pevtsov2007} find SCBs to have a stable mono-polarity in the corresponding photospheric magnetic field yet do not present any data in the associated magnetic field strength.. Without an {\it a priori} bias, this research looks to find statistically significant magnetic substructure underlying the \Ha\ SCBs to establish its conjectured unexpressed unipolar nature. These ambiguities in describing the physical nature of SCBs motivates us to use a consistent technique to reanalyze previously studied SCB events.  

Utilizing the automated detection and tracking techniques developed by~\citet{Kirk2011}, we reanalyze nine previously studied eruption events as well as two new cases where SCBs are found without a flare but with a filament liftoff. Examining the data using a consistent method from these 11 events assists in creating a unified picture of SCBs. We will address three questions. Is there a relationship between any physical properties of the host flare ribbons and SCBs (\emph{e.g.} intensity, velocity, class, timing, number of SCBs)? Is there a consistently measurable photospheric magnetic field component to SCBs? Can a potential coronal field model of SCBs elucidate the other SCB measurements? 

Section~\ref{s:data} describes the events and data incorporated into this study. Section~\ref{s:analysis} briefly explains our methods of feature detection and data assimilation. Section~\ref{s:results} presents the physical characteristics of flares and associated SCBs along with potential field source surface modeling of SCBs. Sections~\ref{s:discussion} and~\ref{s:conclusions} discuss the physical results of the data and the conclusions we draw from them.

 
\section{Events Examined}
     \label{s:data} 
     
     We selected 11 chromospheric ribbon flares to analyze the characteristics of SCBs, which are listed in Table~\ref{t:events}. These events had a favorable viewing geometry with the entire flaring region visible on the solar disk as well as full temporal coverage from the ground-based \Ha\ telescope. Section~\ref{s:isoon} further discusses the details of \Ha\ observations. The photospheric magnetic field of each event was gleaned from cotemporal magnetograms. Section~\ref{s:magneto} addresses the specifics of the treatment of the magnetic field data.  The {\it Geostationary Operational Environmental Satellite} (GOES) measured flare class of each event varied greatly, ranging from no detectable soft X-ray signature in two events, to as large as an X10.0 flare. Corresponding coronagraph images were analyzed for each event to visually search for a CME associated with the flare eruption. Data about the associated CMEs was ascertained from an automated characterization software package and is further discussed in Section~\ref{s:CME_info}. Both the GOES class and the existence of a visual CME also are listed in Table~\ref{t:events}. In this work, we are using the term ``flare eruption'' to include both the overall evolution of the \Ha\ flare ribbons as well as the impulsive appearance SCBs located outside of the flare ribbons.

\begin{table}  
  \caption{The events used in this work to investigate the automated identification and tracking of SCBs and flare ribbons. The time listed is the start time of the flare or filament eruption in \Ha. The heliographic Stonyhurst (HGS) coordinates of the approximate centroid of each event is listed for reference. Nine of the eleven events were previously identified by \citet{Bala2005}, \citet{Bala2006}, \citet{Kirk2012b}, \citet{Kirk2012a}, or \citet{Kirk2014} labeled {\bf Ba}, {\bf Bb}, {\bf Ka}, {\bf Kb} or {\bf Kc} respectively, while two are new in this study. The data sources utilized for each event are abbreviated as: I (for ISOON), L (for Large Angle and Spectrographic Coronagraph (LASCO)), G (for GOES), M (for Michelson Doppler Imager (MDI)), H (for Helioseismic and Magnetic Imager (HMI)), and C (for Computer Aided CME Tracking (CACTus)).}
\begin{tabular}{llllllc}
\dhline

Event & Time & HGS & GOES & Visual & Data Sources & Previously  \\ 
Date  & UT & Coordinates & Class & CME & & Studied  \\
\hline
19 Dec. 2002 & 21:34 & $9^{\circ}$W, $15^{\circ}$N & M2.7 & yes &  I, L, G, M, C & Ba and Kb  \\
6 Mar. 2003 & 15:08 & $0^{\circ}$W, $27^{\circ}$N & None & no & I, L, G, M & no  \\
9 Mar. 2003 & 15:18 & $1^{\circ}$E, $35^{\circ}$N & B6.6 & yes & I, L, G, M, C & Bb  \\
11 Jun. 2003 & 17:27 & $23^{\circ}$E, $16^{\circ}$S & M1.8 & no & I, L, G, M & Bb  \\
29 Oct. 2003 & 20:37 & $9^{\circ}$W, $19^{\circ}$S & X10.0 & yes & I, L, G, M, C & Bb  \\
9 Nov. 2004 & 16:59 & $51^{\circ}$W, $8^{\circ}$N & M8.9  & yes & I, L, G, M, C & Bb and Ka  \\
6 May 2005 & 16:03  & $28^{\circ}$E, $9^{\circ}$S & C8.5 & yes & I, L, G, M, C & Bb and Ka  \\
13 May 2005 & 16:13  & $11^{\circ}$E, $12^{\circ}$N & M8.0 & yes & I, L, G, M, C & Ka  \\
6 Dec. 2006 & 18:29 & $63^{\circ}$E, $6^{\circ}$S & X6.5  & yes & I, L, G, M, C & Kb \\
6 Nov. 2010 & 15:30 & $58^{\circ}$E, $19^{\circ}$S  & M5.4 & yes & I, L, G, M, H, C & Kc \\
30 Nov. 2010 & 17:35  & $39^{\circ}$E, $15^{\circ}$N & None & yes & I, L, G, M, H, C & no \\
\hline
\end{tabular}
\label{t:events}
\end{table} 

Of the 11 flares selected for this study, all but two had been previously examined by \citet{Bala2005}, \citet{Bala2006}, \citet{Kirk2012b}, \citet{Kirk2012a}, or \citet{Kirk2014}. This reanalysis of events will provide two benefits. First, all events are analyzed with the same automated suite of software. This makes the results from each event directly comparable to each other without needing to account for biases within the software or analyst. Second, a comprehensive investigation of photospheric magnetic fields has not been completed for any of the events being studied. This investigation will compile magnetic field data of SCBs, filling a gap in our knowledge of these phenomena. 

\subsection{ISOON \Ha}
	\label{s:isoon}	
	This study examines chromospheric flares and their associated SCBs with \Ha\ (6562.8~\AA) images recorded by the {\it Improved Solar Observing Optical Network} \citep[ISOON:][]{Neidig1998} prototype telescope. ISOON is a ground-based, semi-automated instrument, imaging the full-disk with 1.1 arcsec pixel resolution at a one-minute cadence. Each $2048 \times 2048$ pixel image is normalized to the quiet Sun and corrected for atmospheric refraction. Immediately subsequent to recording the \Ha\ line center images, ISOON also takes two off band images in the red and blue wings of the line at $\pm0.4$~\AA. These wing images are then translated into a Doppler velocity measurement using a Doppler subtraction technique and assuming a consistent and symmetric \Ha\ line profile. This assumption is valid as long as the \Ha\ line remains in absorption which is violated in the core of flares~\citep{Kirk_PhD}.
	
	To fully capture the rise and decay of the flare, images were extracted from the archive beginning approximately an hour before the beginning of the flare eruption and extending an hour after the flare decayed back to pre-flare intensity levels in \Ha. This yielded a data cube with between 400 and 1600 images for each event
	
\subsection{Magnetograms}
	\label{s:magneto}
	
	Photospheric magnetic field measurements are made with the {\it Michelson Doppler Imager}~\citep[MDI:][]{Scherrer1995} which is a space-based polarimeter on the {\it Solar and Heliospheric Observatory} (SOHO) satellite. MDI images the \NiI\ line at 6768~\AA\ every 96 minutes on a $1024 \times 1024$ pixel CCD with a spatial mapping of 2~arcsec {\it per} pixel using a pair of tunable Michelson interferometers. 
	
	For the two events in 2010, high-resolution photospheric magnetic field measurements are available using the {\it Helioseismic and Magnetic Imager}~\citep[HMI:][]{Scherrer2011}. HMI has significantly higher spatial and temporal resolution than MDI. HMI uses a photospheric \FeI\ line at 6173.3~\AA\ and images the Sun on a $4096 \times 4096$ pixel CCD with a spatial scale of 0.6 arcsec {\it per} pixel. Images are recorded at a cadence of 45 seconds and have a 10 G precision. We processes the raw data into their vector components using the HMI Vector Magnetic Field Pipeline~\citep[described by][]{Hoeksema2014} and end up with six images temporally spanning each event. We select the image nearest to the timing of the peak \Ha\ emission of each SCB for this analysis. 

\subsection{CME Data}
	\label{s:CME_info}

Coronagraphic images from the SOHO {\it Large Angle and Spectrographic Coronagraph}~\citep[LASCO ][]{Brueckner1995} C2 instrument are used to visually identify CMEs emerging from the identified \Ha\ events. LASCO is a white light coronagraph, which images the corona from about 1.5 to 6 solar radii. The Computer Aided CME Tracking (CACTus) package is an autonomous software package that detects and characterizes emerging CMEs in LASCO~\citep{Robbrecht2009}.  CACTus detects erupting CMEs and records the central position angle, angular width, and makes a velocity estimation for each CME.


\section{Analysis Techniques}
     \label{s:analysis} 
     
     Each of the flares considered in this study demonstrate several similar physical characteristics. Bright flare ribbons materialize from the active region, separate from each other, undergo an exponential decay in their luminosity, and evolve their topology. Concurrent to the onset of the \Ha\ flare, several types of compact brightenings are observed in the flaring region.  Through careful filtering, we select only those brightenings that exhibit the characteristics of SCBs described in~\citet{Bala2005}, \citet{Bala2006}, and \citet{Kirk2012b}. Next, we employ a Lagrangian approach to identify and trace resolvable subsections of the flare ribbons and SCBs as they appear, disappear, and evolve throughout the eruption. This process was initially developed and documented by \citet{Kirk2011}. These identification processes are designed to analyze specifically the \Ha\ images from ISOON and are described in Section~\ref{s:tracking}. 
     
     By definition, SCBs are sequential in nature~\citep{Bala2005}. Thus, one SCB does not itself characterize the evolving flare environment. Section~\ref{s:ffit} describes the semi-autonomous forward-fitting algorithm developed to distinguish populations of SCBs and determine their collective propagation velocities. 
     
     Subsequent to the identification of SCBs in \Ha, the spatial locations of these points are overlaid on complementary magnetograms from MDI. A potential-field-source-surface (PFSS) model~\citep{Schrijver2003} is then applied to extrapolate the chromospheric and coronal magnetic field lines originating from these locations. A description of the PFSS model used is in Section~\ref{s:pfss}.
     
\subsection{Tracking Algorithm}
	\label{s:tracking}

There are two steps needed to extract individual kernels from \Ha\ images: detection and tracking. A kernel is defined in this work to be a small locus of pixels that have increased intensity, which can be isolated from other pixels in the immediate vicinity. The detection algorithm identifies preliminary bright kernels in a set of images by eliminating pixels that are dimmer than a specific intensity (1.35 times background intensities for flares and 1.2 for SCBs) as empirically determined by~\citet{Kirk2011}. Both low and high spatial bandpass filters are applied to isolate features and suppress noise.  Each preliminary kernel does not have any predetermined size or shape, but does have a local maximum and is isolated from its nearest neighbors by at least one ``dark'' pixel. Next, properties of the candidate kernels are calculated: integrated intensity, radius\footnote{The calculation of a kernel ``radius'' is more accurately a ``radius of gyration'' and is calculated by finding the mean intensity weighted radius from each pixel to the kernel axis of rotation. See~\citet{Crocker1996} for further details.}, and eccentricity. The candidates are then filtered by size, shape, and intensity to eliminate unwanted features such as broad regions brightening in concert or single pixel detections of noise. For a complete discussion of the process of kernel detection and filtering, see~\citet{Kirk2011}.

The second step in flare and SCB kernel extraction process is linking time-resolved kernels between individual frames in \Ha. These trajectories allow us to isolate single kernels and follow their evolution through time. We employ a diffusion-based algorithm to statistically associate similar kernels between images. This tracking technique was initially developed by~\citet{Crocker1996} and subsequently modified by~\citet{Kirk2011} for tracking solar features.  This statistical approach maximizes the probability that a single particle with classical Brownian motion will diffuse within a set range of distances in a given segment of time. The diffusion probability is generalized for a system of any number of non-interacting particles. This probabilistic approach to tracking yields trajectories for all identified kernels at once. Once initial trajectories are identified, a filter is applied to eliminate weak detections lasting less than a minimum number frames~\citep[see][for a complete description]{Kirk2011}. This temporal filter eliminates off-ribbon flare detections which are associated with the eruption but do not characterize the evolution of the flare ribbons and SCB kernels that have ambiguous definition ({\it i.e.} they could be flare kernels). 

The end result is a set of flare and SCB kernels that individually appear stochastic, yet fully represent and characterize the evolving flare region as an ensemble and allow us to extract quantities of interest such as location, velocity, and intensity of subsections of the flare ribbons and individual SCBs.

\subsection{Forward Fitting}
	\label{s:ffit}

To identify groups of SCBs and characterize their propagation, a slope is required to be fit to the time-distance SCB data set. A simple regression analysis is impossible in this situation because multiple populations of SCBs may exist in the same data set. A more complicated mixture modeling technique is also unreliable because there is a relative small number of SCBs detected which leads to large statistical uncertainties and ambiguous group identification. Therefore, we employ a supervised, iterative, forward-fitting technique, similar to a linear discriminate analysis. This method requires the user to first, visually determine the number of groups to be fitted, and second, identify the approximate location of each group. The forward-fitting routine then searches all linear combinations of features for the next ``best point" to include in each of the groups, which minimizes the $\chi^2$-value to the identified group.  This process is then repeated with each of the groups that now include points from the previous iteration. Repeating this method over all the points in the set produces an ordered set of points that when sequentially fit, have an increasing $\chi^2$-value. 

This iteration allows us to view the set of SCB data in $\chi^2$-space by characterizing how a selected group variance changes as each point in the set is added. Selecting the groups in the $\chi^2$-curve has the effect of identifying where the variance within a group begins to increase dramatically. The derivative of the $\chi^2$ distribution is then taken and the point in which this curve increases to beyond one standard deviation is selected to divide the population of one group from the rest of the set.  Running this routine several times, each time selecting the same groups, minimizes the effect of the user and provides an estimate of the error associated with the group identification.

This method has two caveats. First, this fitting method relies on the detections having Poissonian noise. This presumption is imprecise, since the detection process of compact brightenings introduces a selection bias. Second, the fitting method assumes that no acceleration occurs in the propagation of SCBs. This is a reasonable first-order approximation from the studies of SCB group propagation by~\citet{Bala2005}.

\subsection{Potential Field Model}
	\label{s:pfss}
	
	Polarimeters are used to infer the photospheric magnetic fields because that is the only place in the solar atmosphere with consistently adequate light intensity to measure polarization. To extrapolate the inferred photospheric magnetic field to the rest of the upper atmosphere, a model is needed. The simplest model is the potential-field-source-surface (PFSS) which assumes a potential atmosphere that is current-free. The challenge for any model is to find the current free scalar potential in a spherical geometry. 

Schrijver and De Rosa (2003) developed an assimilation PFSS model to extrapolate a current-free potential within a spherical volume between $r = 1~R_\odot$ and $r = 2.5~R_\odot$ given an initial photospheric magnetic field.\footnote{DeRosa's PFSS modeling software package is currently available at {\tt www.lmsal.com/$\sim$derosa/pfsspack/}} The model derives a unique solution for a given domain and boundary conditions both at the top ($r = 2.5~R_\odot$) and bottom ($r = 1~R_\odot$). At the upper boundary condition the field is assumed to be purely radial. The lower boundary condition field is derived from an evolving surface-flux transport model~\citep{Schrijver2001}. Synoptic magnetograms from MDI are used to anchor the model in the photosphere. The flux transport model then assimilates the photospheric field through the entire domain by advecting the flux stepwise vertically across the full solar surface.  It empirically determines differential rotation, meridional flow, and convective dispersal profiles from the input data, using a non-linear algorithm to account for fragmentation and collision of flux. The result of the PFSS model is a 3D projection of the magnetic field in which field lines can be drawn. This model is limited to locations where the field is potential, which is most accurate in the quiet Sun and least accurate in flaring regions where a significant current sheet is generated. 


\section{Physical Results of Algorithm Application}
	\label{s:results}
	
	Applying the detection and tracking algorithm to the 11 flaring events, we identified a total of over 10,000 discernible flare kernels and over 4,000 discrete SCBs in \Ha\ images. Sequential chromospheric brightenings, although related to the erupting flare ribbons, are distinctly different from the flare kernels. The differences between these two types of brightening in the chromosphere is outlined in Table~\ref{t:kernelresults}. Individual SCBs are much more fleeting, smaller, and dimmer than the flare ribbons. A typical SCB lasts less than 10 frames (corresponding to 10 minutes) above the detection thresholds in ISOON \Ha, while flare kernels last on average over 120 minutes.  The diameter of the smallest resolvable kernel along the flare ribbon is approximately 6400~km as compared to SCBs that are resolved down to a 1600~km diameter. A typical SCB has a peak intensity of 1.2 -- 2.5 times brighter than the average background intensity level. In contrast, flare ribbons often brighten more than an order of magnitude above the pre-flare brightness.
	
\begin{table}     
  \caption{General physical characteristics of individual flare and SCB kernels. ``Ensemble Motion'' refers to the motion of an individual kernel as compared to its nearest neighbors over the kernel lifetime. }
\begin{tabular}{lcccc}
\dhline
Kernel & Minimum & Peak Intensity & Average & Ensemble Motion \\
Type & Diameter & Increase & Lifetime &  \\
\hline
Flare & 6.4~Mm & $\ge 1000$\% & $\approx 120$~min & Directional Consistency \\
SCB & 1.6~Mm& $\le 250$\% & $\approx 10$~min & Random Walk \\
\hline
 \end{tabular}
  
\label{t:kernelresults}
\end{table}	

When the individual tracks of SCB kernels are examined, they do not show any progressive motion. The centroid of an SCB kernel randomly ``walks'' around within about six pixels of its starting location for the duration of the trajectory. Although the ensemble of SCBs demonstrate a sequence of point brightenings, giving the appearance of a progressive traveling disturbance, the bright emission of an individually measured SCB does not follow the disturbance and remains in the same location. Similar to a wave, the medium in which SCBs are measured remains laterally undisplaced with the apparent propagation of the brightenings. This result confirms the findings of \citet{Bala2005} and \citet{Kirk2012a}.
	
In addition to these initial results of detection, further insight into the physical processes of the eruption can be gleaned with closer analysis. These are presented in the following two sections: those determined from \Ha\ (Section~\ref{s:isoonresults}) and the results of overlaying the detections on corresponding magnetograms (Section~\ref{s:bfieldresults}).

\subsection{Findings in \Ha}
	\label{s:isoonresults}
	
\begin{figure}    
   \centerline{\includegraphics[width=1.0\textwidth,clip=0,angle=0]{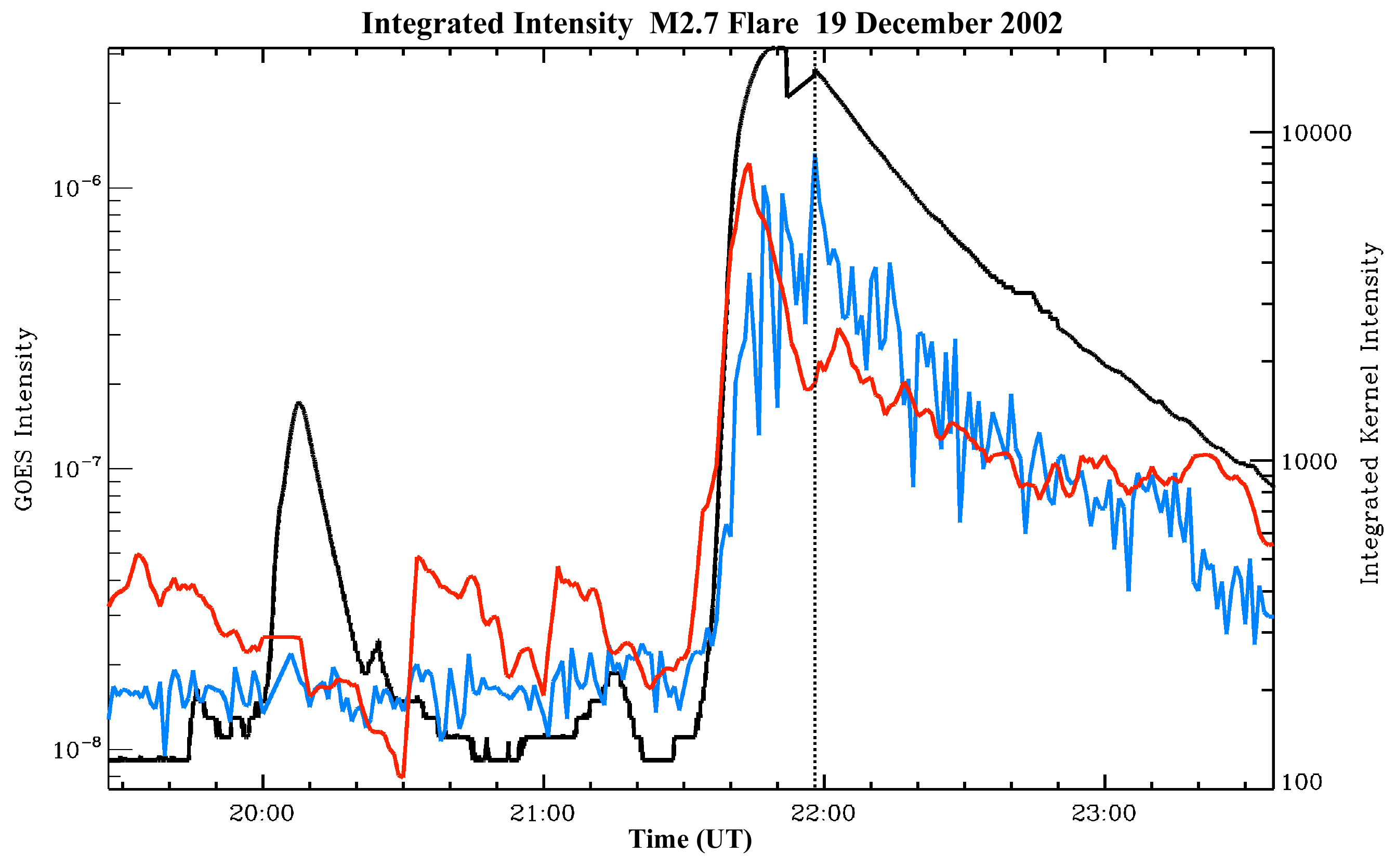}}
  
              \caption{Time evolution of the 19 December 2002 event. The dashed line marks the peak \Ha\ flare intensity. The blue line is the \Ha\ flare kernel intensities integrated at each time step. The redish line plots the SCB kernel intensities (multiplied by a factor of 10 for display purposes) integrated at each time step. Plotted for reference in black is the GOES 1.0 -- 8.0~\AA\ intensity curve.}
   \label{20021219FlareCurve}
   \end{figure}

It is simple to recover the total \Ha\ flare intensity curve since the flare kernels are just fragments of the entire flare. Summing each flare kernel results in an intensity evolution in \Ha\ that is comparable as to the X-ray intensity, the impulsive rise, and decay phase of the flare measured with GOES. Summing each SCB kernel results in an intensity curve that has characteristics comparable to the \Ha\ flare intensity and X-ray intensity curves, yet is distinct.  Figure~\ref{20021219FlareCurve} plots this aggregated intensity of the entire population of flare kernels and SCB kernels {\it versus} time as well as the GOES X-ray intensity in the 19 December 2002 event. In this case, the SCB aggregate intensities peak well before the flare does. The peak of the aggregate SCB intensity curve occurred before or concurrent with the peak \Ha\ flare intensity in 73\% (representing 8 in 11) of the cases studied (Table~\ref{t:scbresults}). 

This analysis of \Ha\ flares yields a couple of interesting results on this well-studied topic. These are discussed in Section~\ref{s:flares}. Separating SCBs from their corresponding flares also allows comparison and contrast between SCBs of differing events and is discussed in Section~\ref{s:scb}.

\subsubsection{Properties of \Ha\ Flares}
\label{s:flares}

\begin{figure}    
   \centerline{\includegraphics[width=0.7\textwidth,clip=0,angle=0]{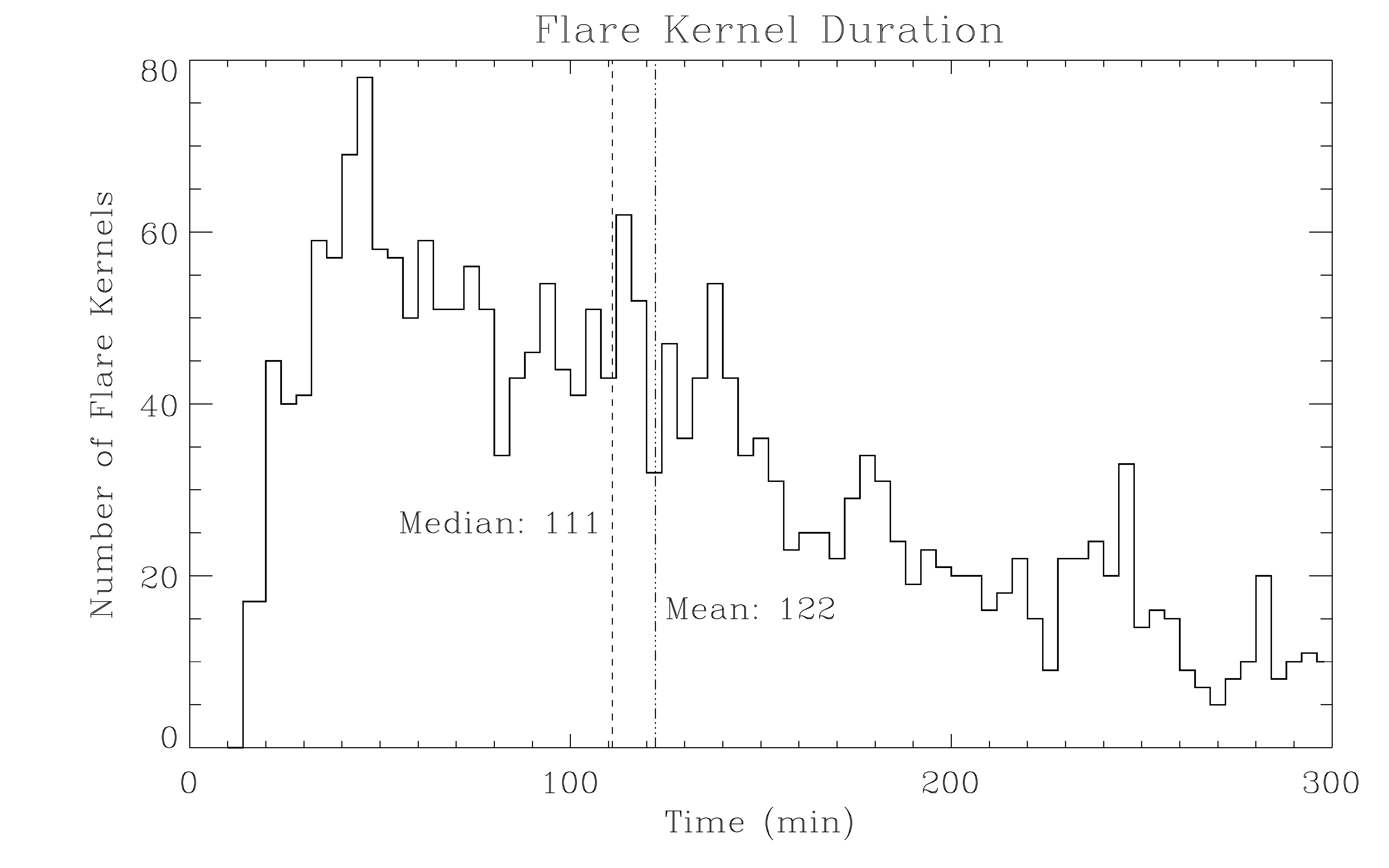}}
  
              \caption{A histogram of the continuance of detected flare kernels in all 11 events. Flare kernels are tracked for a mean duration of 122 minutes (long and short-dashed line) and a median duration of 111 minutes (dashed line).}
   \label{FWHM_Flare_kernel}
   \end{figure}

A result of parsing flare ribbons into kernels is that the temporal integrity of flare substructure is exposed. Distinct \Ha\ flare kernels are temporally robust, lasting on average $\approx122$ minutes each and had a most probable duration of 46 minutes (Figure~\ref{FWHM_Flare_kernel}). This average duration of a single kernel is shorter than the lifetime of the total \Ha\ flare (from the impulsive phase, through peak intensity, and return to pre-flare brightness). Thus, the number of detectable kernels declines as the flare's intensity decays from its peak, implying that there are fewer resolvable components in the flare ribbons as the flare evolves.  This change in detectable kernels implies that the majority of kernels cannot be tracked from pre-flare to post-flare, suggesting a dynamic substructure to the flare ribbons when bright points appear and disappear as the flare erupts. 

\citet{Kirk2012a} also found that individual subsections of flare ribbons can be tracked and are observed to appear and disappear as the underlying flare ribbons evolve. Examining individual kernel structure suggests there is substructure within a flare ribbon whose elements impulsively brighten and dim within the integrated \Ha\ flare intensity curve. These results support the premise that flare ribbons are composed of several magnetic field lines successively reconnecting and depositing energy in the chromosphere~\citep[\emph{e.g.}][]{Priest2002}. There is no evidence to claim an individual flare kernel is directly tracking one of these reconnections or a single post-flare loop footpoint. Thus, within a tracked flare kernel, multiple coronal reconnection events may be superimposed to produce the observed duration of a single flare kernel. Contemporary studies of flare ribbon substructure in the transition region have resolved unique features with a spatial scale of $\approx 2$~Mm and duration of 2\,--\,3 minutes~\citep{Brannon2015}. 

\subsubsection{Properties of SCBs}
\label{s:scb}
\begin{figure}    
   \centerline{\includegraphics[width=0.7\textwidth,clip=0,angle=0]{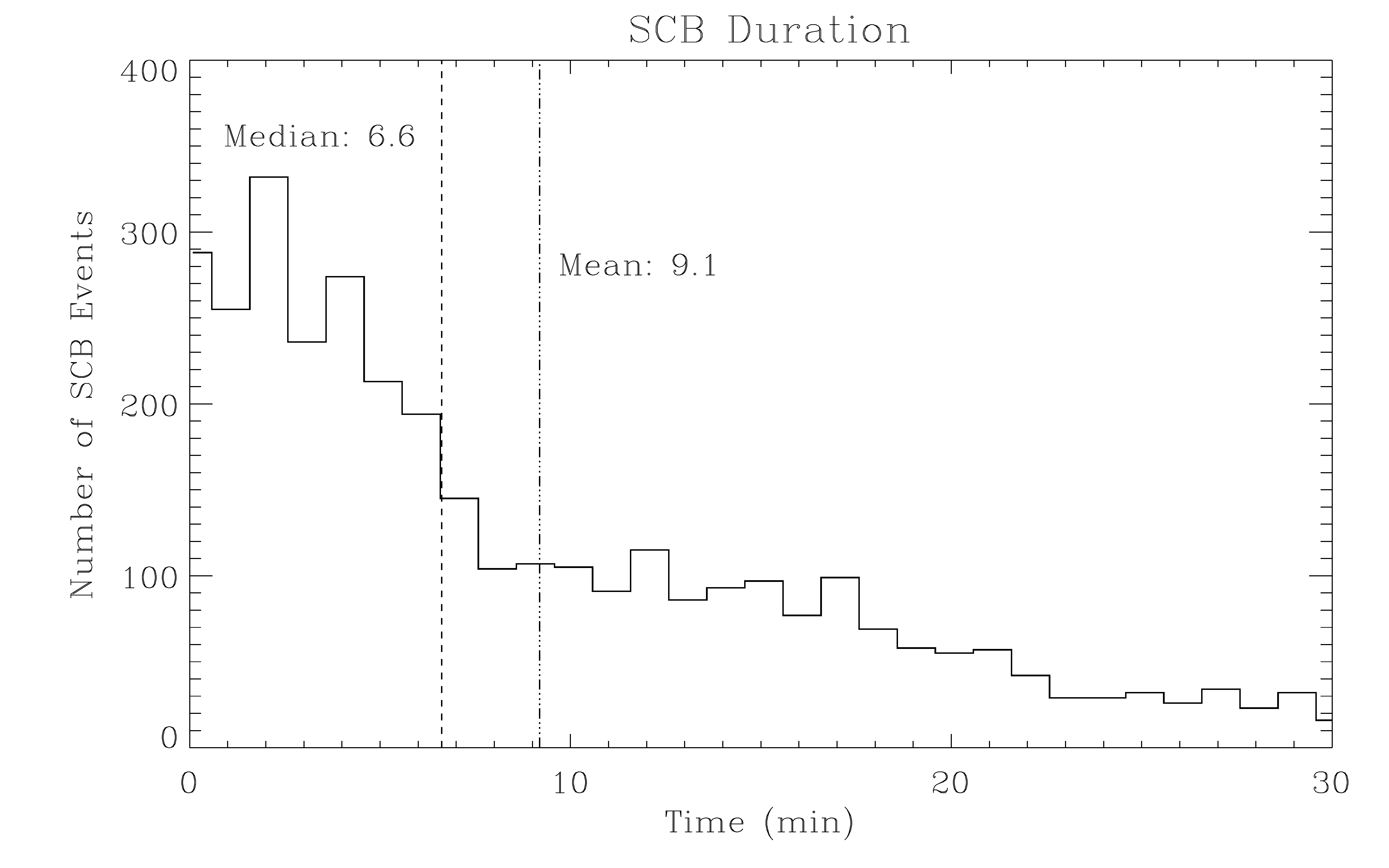}}
  
              \caption{A histogram of the duration (FWHM) of SCB detections in all 11 events. The SCBs have a measured mean duration of 9.1 minutes (long and short-dashed line) and a median duration of 6.6 minutes (dashed line).}
   \label{FWHM_SCB_kernel}
   \end{figure}

The number of SCBs identified in a given event varied greatly from as few as 48 to as many as 1335 (Table~\ref{t:scbresults}). Individual SCBs have light curves that are typically impulsive with a sharp peak -- followed by a return to background intensity with a median duration of 6.6 minutes and a most likely duration of 2.1 minutes (Figure~\ref{FWHM_SCB_kernel}). 

SCBs are more closely related to the impulsive phase of the \Ha\ flare than any other part. However, many SCBs precede even the earliest onset of the associated flare. Figure~\ref{20021219FlareCurve} shows a typical example where SCBs brighten in a relatively short period of time before quickly decaying to an enhanced but consistent intensity. In 10 out of 11 flares examined, SCBs begin brightening in the impulsive phase of the \Ha\ flare, with a peak occurring between 30 minutes and concurrent with the \Ha\ flare peak, returning to an idle intensity in the early decay phase, about 30 minutes after the peak (Table~\ref{t:scbresults}). In contrast, the \Ha\ flare intensity curve remains above quiescent levels for several hours. 

\begin{table}
  \caption{Derived physical measurements for SCBs in each event. Propagation velocities without stated uncertainties were marginal with a high amount of marginal SCB detections. }
\begin{tabular}{llllll}     
\dhline
Event & Number & SCB Velocity & SCB Timing & Mean MDI & CME Speed\\
Date & of SCBs & (\kms) & {\it vs.}  Eruption & $\vec{B}$ (G) & (\kms) \\
\hline
19 Dec. 2002  	& 562 	&	$84.8\pm 9.7$					& Before 		& 1.8 	& 697 \\
			&		&	 $260\pm 19.4$					&			&		&	\\
6 Mar. 2003 	& 684 	&	$220\pm143$					& Coincident 	& -2.5 	& -- \\
9 Mar. 2003 	& 57 		&	$73.9\pm 10.9$					& After 		& -0.5 	& 187 \\
			&		&	 -$8.5\pm 3.6$					&			&		&	\\
11 Jun. 2003 	& 281 	&	$95.7\pm 40.0$					& After 		& 1.7 	& -- \\
			&		&	 -$1684\pm 1914 $					&			&		&	\\
29 Oct. 2003 	& 1335 	&	$460$						& Coincident 	& -6.7 	& 2029 \\
			&		&	 $2423$						&			&		&	\\
9 Nov. 2004	& 302 	&	-- 							& Before 		& 5.9 	& 1562 \\
6 May 2005 	& 171 	&	$63.0\pm 15.8$					& Before 		& -4.7 	& 917\\
13 May 2005 	& 154 	&	$36.3\pm 7.2$					& Coincident 	&12.2 	& 553 \\
			&		&	 $153\pm 55.8$					&			&		&	\\
6 Dec. 2006 	& 291 	&	$851$						& Before 		& 0.5 	& 984 \\
6 Nov. 2010	& 210	&	$65.4\pm 4.8$					& After		&-6.2 	& 206 \\
			&		&	 $465\pm 206$					&			&		&	\\
30 Nov. 2010 	& 48 		& 	$51.0\pm 4.9$ 					& Before 		& 3.6 	& 282 \\
\hline
 \end{tabular}  
\label{t:scbresults}
\end{table}

\begin{figure}
   \centerline{\includegraphics[width=0.6\textwidth,clip=0,angle=0]{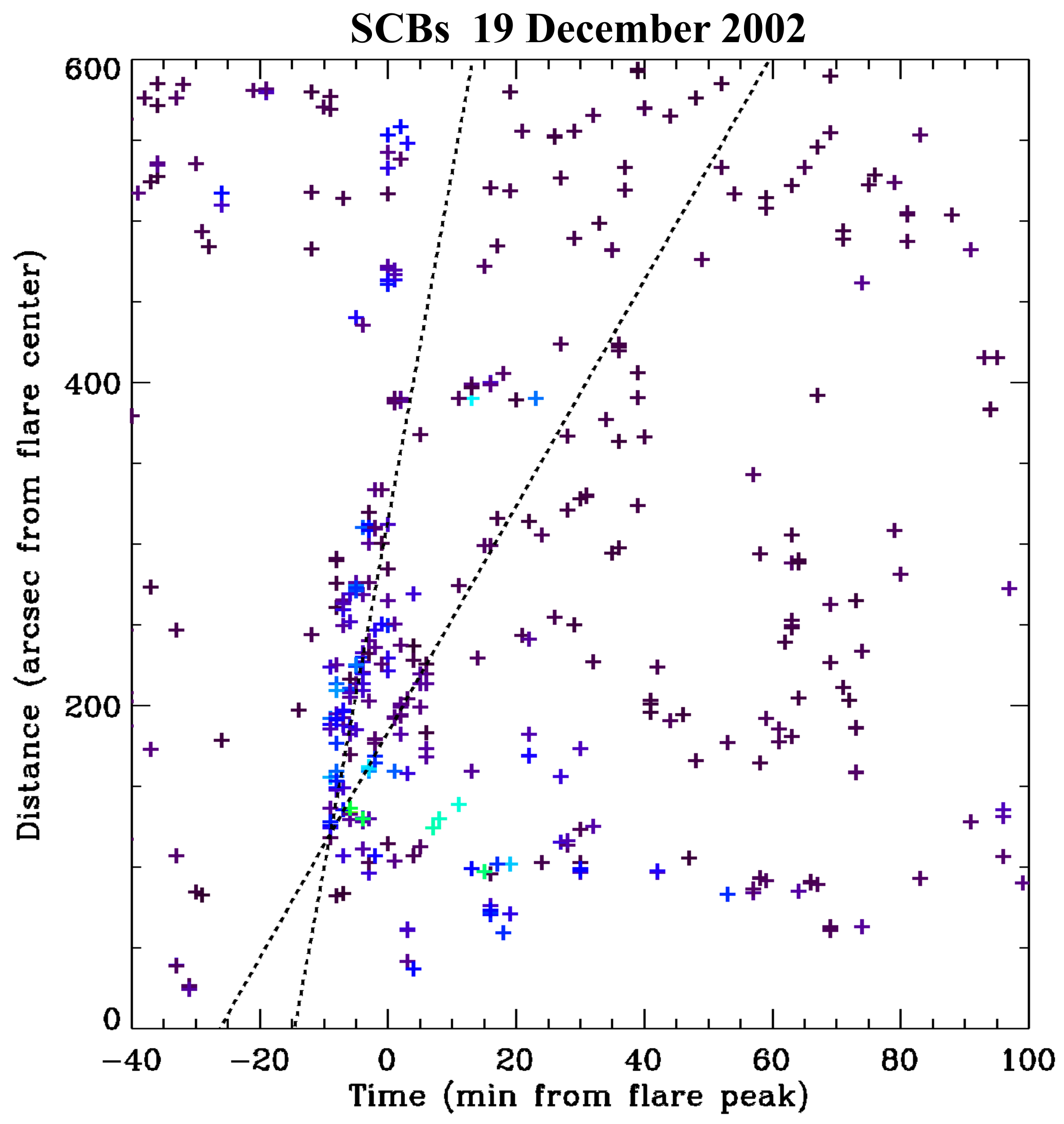}}
 
              \caption{The distance at which the bright point occurs {\it versus} time from the \Ha\ flare peak for SCBs during the 19 December 2002 event. The color of each plotted point is representative of its relative intensity: the dimmest are purple, higher intensity detections are blue, and the highest intensity detections are green. The black dashed lines show two weighted regression fits: $84.8\pm9.7$~\kms, $260\pm19.4$~\kms.}
   \label{20021219DistTime}
   \end{figure}

SCBs tend to cluster together in time-distance plots in all events. In Figure~\ref{20021219DistTime} the shade of the mark (from violet to green) corresponds to the intensity of the SCB measured. The closer the color of the mark is to green, the brighter the SCB. Generally, the brighter SCBs are spatially closer to the flare center and temporally closer to the \Ha\ flare peak intensity. This intensity correlation is weak and is more closely related to distance rather than time of brightening. 

Applying the forward-fitting technique to these data sets yields three propagation speeds: slow propagation, fast propagation, and surge propagation. The dashed lines in Figures~\ref{20021219DistTime} visually show the groups identified and Table~\ref{t:scbresults} lists the results. Slow propagation of less than or equal to 100 \kms\ was measured in 64\% (7 of 11) of the events. Their velocities ranged from 36.3 to 84.8 \kms\ (a tenuous measurement of velocity, $95.7\pm40$ \kms, occurred in the 11 June 2003 event but we exclude it in this discussion because of the large errors associated). A fast propagation was measured in 64\% (7 of 11) of the events. More than half (6 of 11) of the events exhibited two types of propagation simultaneously. Just two events demonstrated fast propagation without a slow group as well.  These fast speeds ranged from 153 to 851 \kms\ and typically had large relative errors associated with them.  Two events (11 June 2003 and 29 October 2003) had a surging propagation groups identified. Both of these measurements were on the order of the coronal Alfv\'{e}n speed, at over 1000 \kms, and had errors larger than the measurement. The 9 May 2003 event was the only event measured to have a negative slow propagation velocity, meaning that SCBs were approaching the flare.
 
\subsection{SCBs in Magnetograms}
	\label{s:bfieldresults}

Previous studies including \citet{Bala2006}, \citet{Pevtsov2007}, and \citet{Kirk2012a} all suggest that SCBs are strongly monopolar based on an empirical model. To extend these findings, photospheric line-of-sight (LOS) magnetic field measurements from MDI and HMI were extracted from the same region of interest identified in ISOON (Figure~\ref{SCBMag}). The locations and spatial footprint of the identified SCBs were overlaid onto the magnetograms and mean magnetic fields calculated for each SCB. The results of magnetic field measurements using MDI are presented in Section~\ref{s:bmeasuredMDI}, which provide the basis for modeling of the magnetic field out of the photosphere (presented in Section~\ref{s:bmodeled}). The measurements of the vector magnetic field for the two events in 2010 are presented in  Section~\ref{s:bmeasuredHMI}.

\subsubsection{MDI Measurements}
\label{s:bmeasuredMDI}

\begin{figure}
   \centering
  
   \includegraphics[width=0.7\textwidth,clip=0,angle=0]{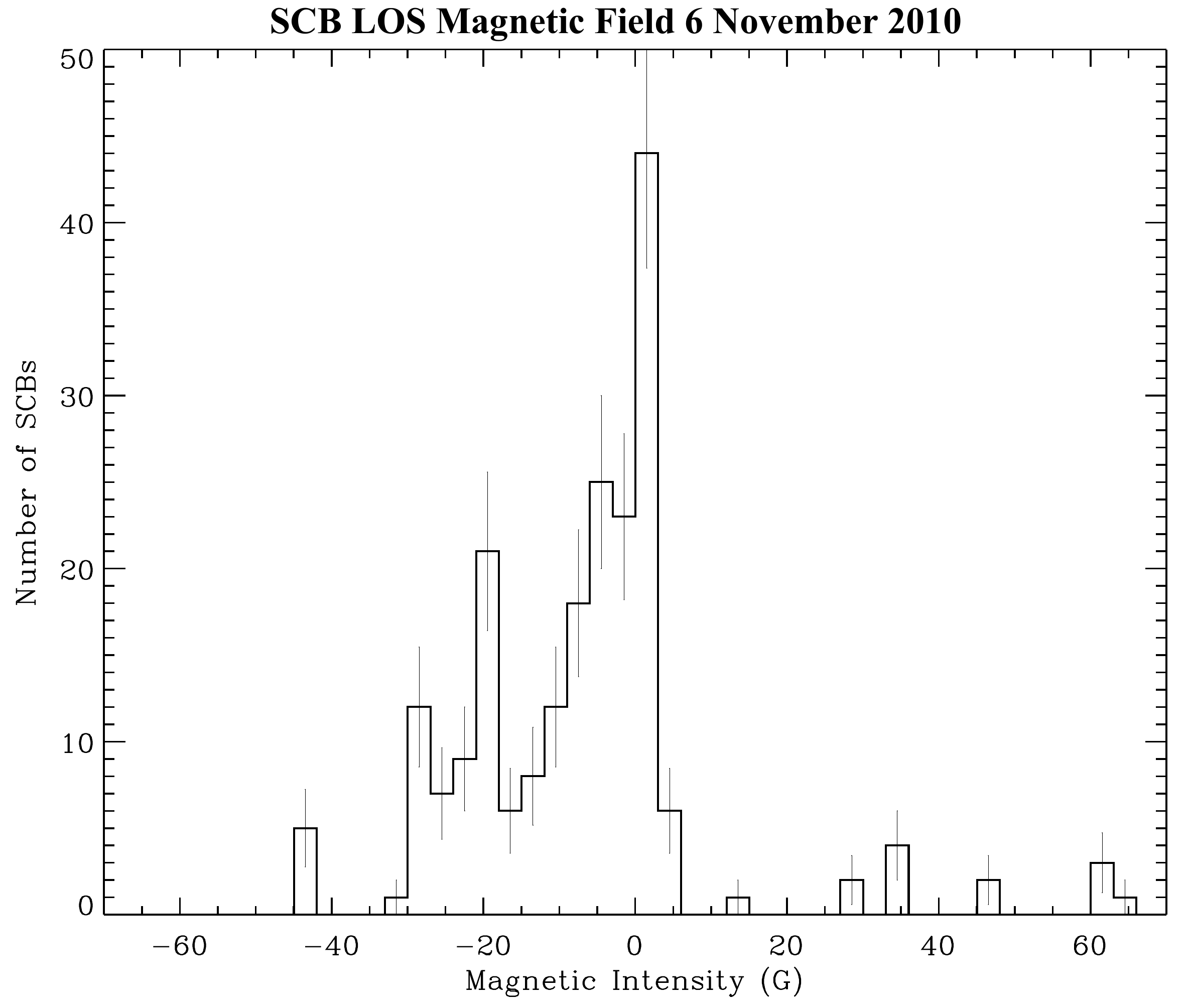}
  
             \caption{The distribution of photospheric LOS magnetic field intensities in MDI beneath SCBs for the 6 November 2010 event with a mean magnetic intensity of -6.2~G and a medan of -5.1~G.}
   \label{20021219BCurve}
   \end{figure}

The majority of the photospheric SCB magnetic field have a range between $\pm 50$ G. A histogram of individual SCBs measured magnetic intensity is plotted for the 6 November event in Figure~\ref{20021219BCurve}. The mean of the magnetic field measurements for all events are also reported in Table~\ref{t:scbresults}, ranging from $-6.7$~G in the 29 October 2003 event to $12.7$~G in the 13 May 2005 event. Magnetic intensity histograms of each event show some substructure, suggesting multiple groups of SCBs within a single flare.  If SCBs did exhibit strong monopolarity, a histogram of magnetic field measurements should have a bifurcated distribution. The population of measurements observed did show a weak bias toward one polarity; however, the most common measurement was always $\pm 0.5$~G around 0~G in every event except 6 May 2005. There is also no apparent correlation between SCB \Ha\ and magnetic intensities.  These findings contradict the empirical model of previous studies in that brightness and location of SCBs are not directly related to magnetic polarity or intensity. These older measurements do show that a typical individual SCB has a distinct non-zero magnetic polarity. 

\subsubsection{HMI Measurements}
\label{s:bmeasuredHMI}

There were two events in 2010 where the SCB detections had cotemperal HMI magnetograms. We further processed these magnetograms into their vector components: $\vec{B}_\phi$ zonal, $\vec{B}_\theta$ meridional, and $\vec{B}_r$ radial  (see Section~\ref{s:magneto}). The events on 6 November and 30 November had 210 and 48 individual SCBs detected respectively, which represents about 6\% of the total number of SCBs detected in all events. Thus these measurements of SCB vector magnetic components in 2010 may not be representative of all SCBs, but do give us some significant insight beyond magnetic field measurements in MDI.  

\begin{figure}
   \centering
  
   \includegraphics[width=0.8\textwidth,clip=0,angle=0]{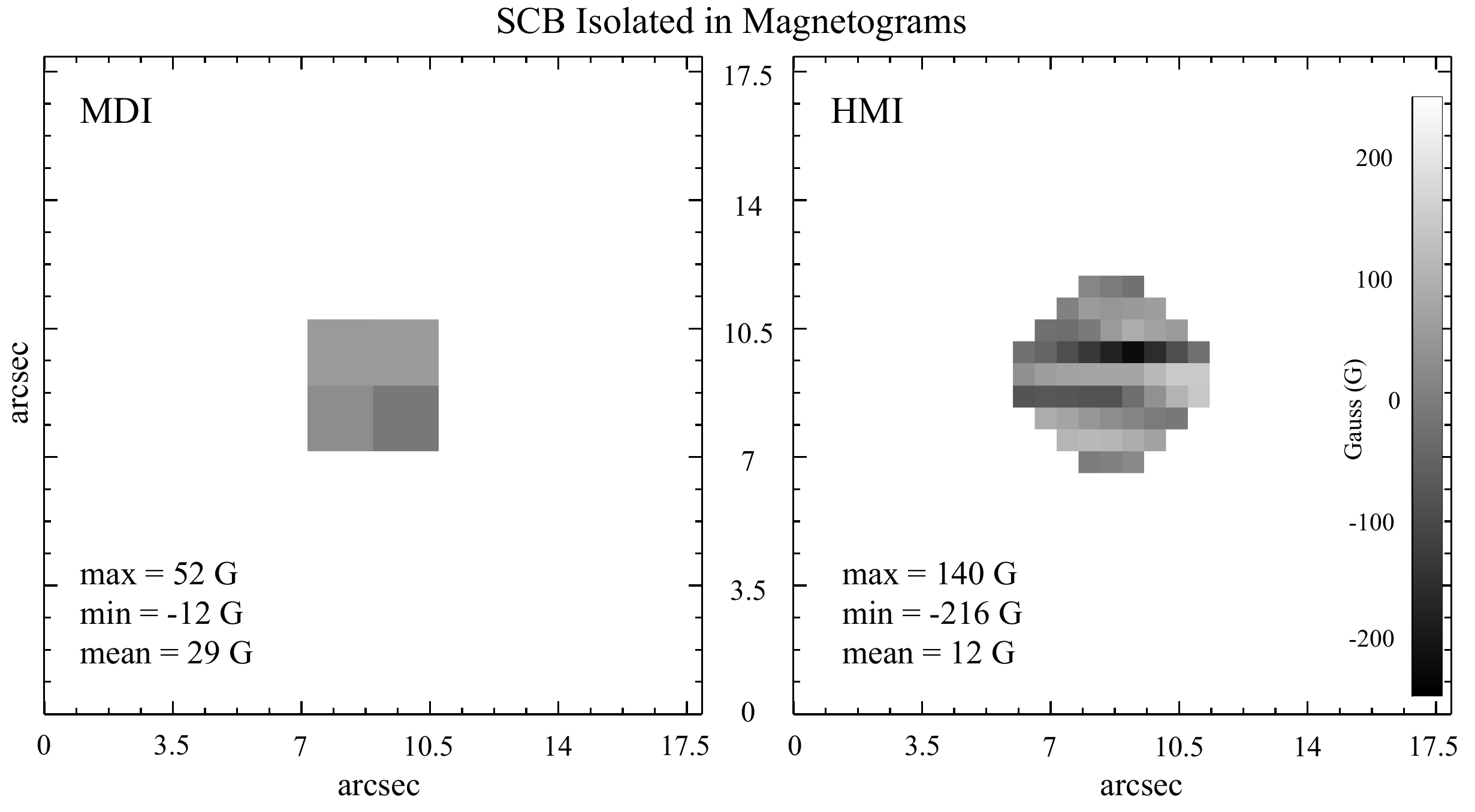}
  
             \caption{An example of a single SCB from the 6 November 2010 event isolated in MDI and HMI. The boundary of the SCB is determined by the flux in \Ha\ and then overlaid on the respective photospheric magnetogram to isolate the magnetic signature of the SCB. Both images are plotted on the same intensity gray scale shown on the right.}
   \label{SCBMag}
   \end{figure}

The power of measuring magnetic fields using the higher resolution of HMI is demonstrated in Figure~\ref{SCBMag}. The pixel scale of MDI is too coarse to include more than a few pixels to measure the magnetic intensity. HMI has superior resolution both in the spatial direction but also in measuring the magnetic flux. MDI is insensitive to the small-scale variations of magnetic flux within the boundary of the SCB. Figure~\ref{SCBMag} graphically shows that there is substructure within an SCB that has gone previously unobserved using MDI. To demonstrate that the distribution of these magnetic elements is not purely statistical noise, we calculate the skewness and kurtosis within each SCB (see discussion in the following paragraphs). The mean magnetic flux of the SCB measured in each of these magnetograms is comparable, but the extrema are significantly different. Using HMI, a small negative magnetic feature is clearly visible completely bounded by the SCB. This feature is not apparent in MDI.   

\begin{figure}
   \centering
  
   \includegraphics[width=0.9\textwidth,clip=0,angle=0]{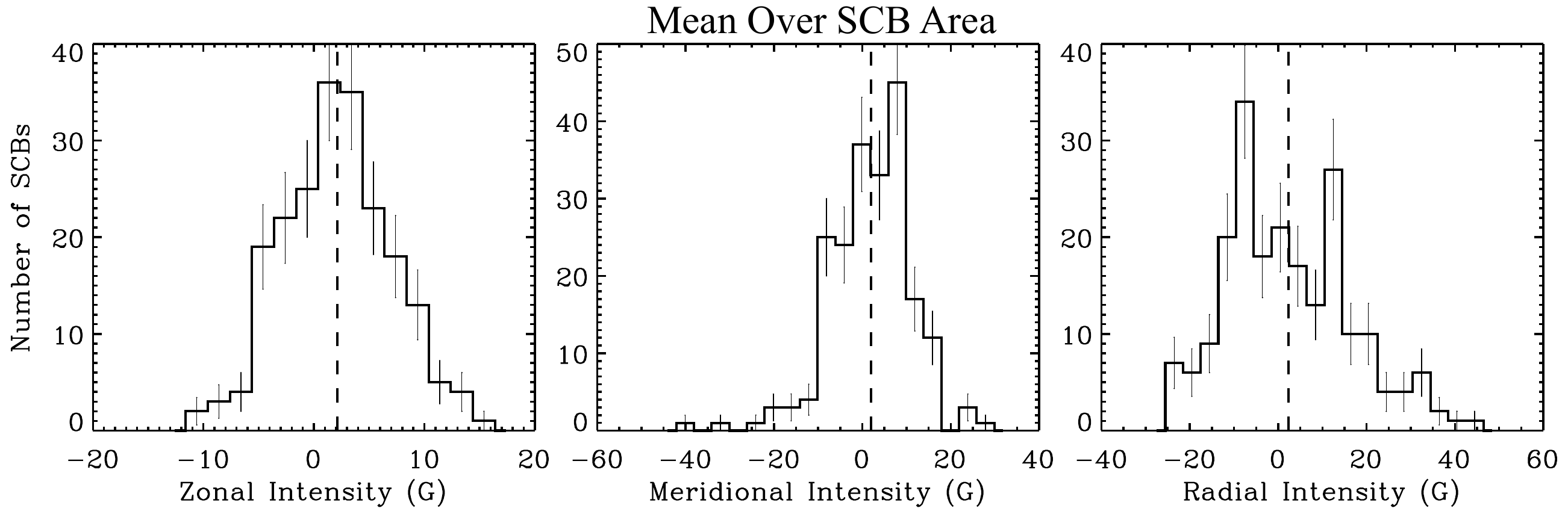}
   \includegraphics[width=0.9\textwidth,clip=0,angle=0]{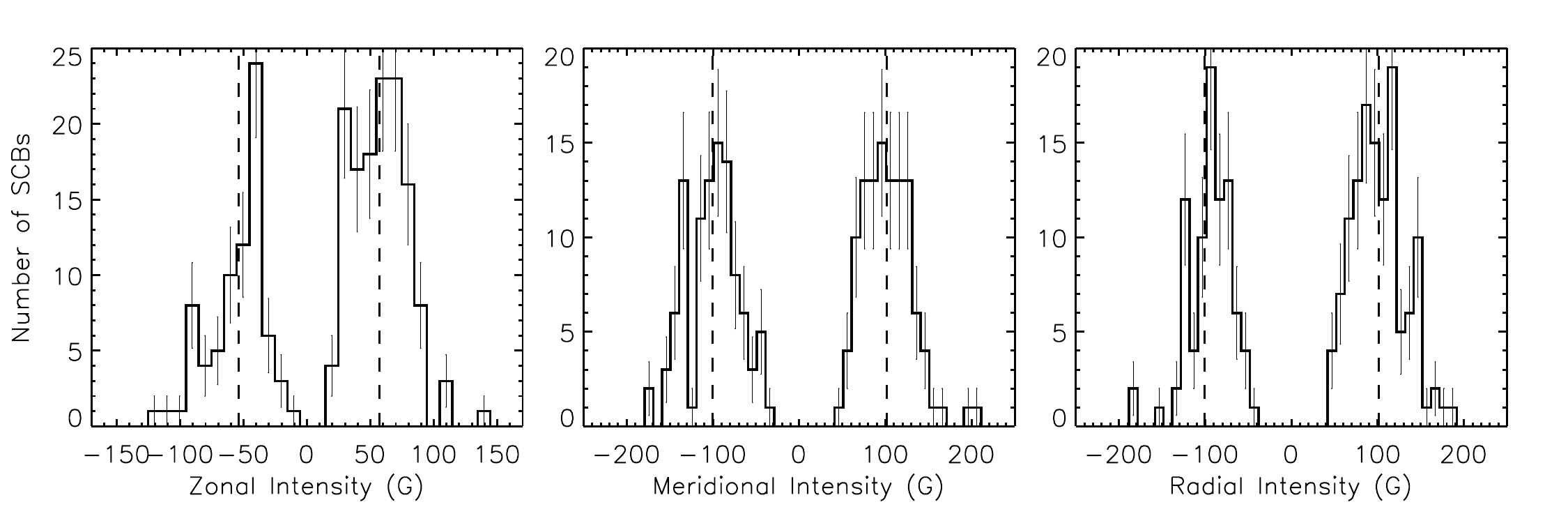}
  
             \caption{The distribution of photospheric vector magnetic field intensities beneath SCBs for the 6 November 2010 event. The dashed lines show the mean intensity of each distribution.  The top row averages the HMI vector components within each SCB: $\langle \vec{B}_{\phi,\textrm{\tiny SCB}} \rangle$, $\langle\vec{B}_{\theta,\textrm{\tiny SCB}}\rangle$, and $\langle\vec{B}_{r,\textrm{\tiny SCB}}\rangle$. The bottom row shows the vector components of the single pixel with the largest magnitude bounded by the SCB: $\vec{B}_{\phi,B_\textrm{\tiny max}}$, $\vec{B}_{\theta,B_\textrm{\tiny max}}$, and $\vec{B}_{r,B_\textrm{\tiny max}}$. The statistics of the extrema measurements are listed in Table~\ref{t:HMIresults}.   }
   \label{20101106BDist}
   \end{figure}

The population of HMI LOS pixels over all SCBs is regularly distributed. The skewness averaged over each HMI measurement is near zero ($\langle \gamma_1 \rangle = 0.03$) indicating that the HMI measurements {\it en masse} are uniformly distributed about their mean value. In other words, in a conglomerate of all SCBs, the positive and negative polarities are equally represented. This does not mean that within a given SCB, the skewness is near zero. The standard deviation of the skewness measurements over all SCBs ($\sigma_{\gamma_1}=0.35$) indicates there are significant numbers of individual SCBs with positive or negative skewness. The near-zero average skewness combined with the average distribution of HMI pixels measured near zero (Figure~\ref{20101106BDist}) means that SCBs {\it en masse} do not have a preferred polarity.

To calculate how likely an extreme value is to occur, we calculate the kurtosis. The excess kurtosis aggregated over each SCB measurement is consistently platykurtic ($\langle \textrm{Kurt} \rangle =-0.63 \pm 0.23$). This means that outliers in the tails of this distribution are less likely to occur because of random statistical noise.  Given the typical values of the mean, skewness, and kurtosis, a magnetic extrema measured within the SCB boundary has a greater significance than its naively calculated standard error. The magnetic extrema within an SCB have values that are typically five standard errors beyond the mean value of the population ($B_\textrm{max}= \langle\vec{B}_{\textrm{\tiny SCB}}\rangle+5.2\langle \textrm{S$_E$} \rangle$).  The calculation of the standard error assumes a Gaussian model, which overestimates the likelihood in the tails of our platykurtic distribution, so the true standard error is likely to be smaller and the extrema significance greater ($> 5.2$). The extrema selected have significantly stronger magnetic flux as compared to the rest of the HMI pixels within the bounds of an SCB.  

Magnetic extreme values measured within an SCB have statistical significance, which makes them a logical choice to isolate from the other pixel values measured.   Given the statistical significance (a typical value $> 5.2$ standard errors above the mean) of the extreme values within each SCB, it naturally follows to infer a physical meaning to the reason why an SCB contains such magnetic outliers. These outliers imply that magnetic footpoints of an SCB in the photosphere is significantly smaller than the corresponding area of the \Ha\ chromospheric brightening.

Figure~\ref{20101106BDist} shows the distributions of the zonal, meridional, and radial magnetic field components under SCBs in the 6 November 2010 event. We report the magnetic field components in two ways. First we calculate the mean magnetic field component magnitude over a given SCB area: $\langle\vec{B}_{\phi,\textrm{\tiny SCB}}\rangle$, $\langle\vec{B}_{\theta,\textrm{\tiny SCB}}\rangle$, and $\langle\vec{B}_{r,\textrm{\tiny SCB}}\rangle$ (Figure~\ref{20101106BDist} top). Then we identify the pixel with the largest magnetic field magnitude, L1 norm, within the SCB area, $B_\textrm{max}=\max(||\vec{B}_{SCB}||)$, and then measure the vector components of that pixel: $\vec{B}_{\phi,B_\textrm{\tiny max}}$, $\vec{B}_{\theta,B_\textrm{\tiny max}}$, and $\vec{B}_{r,B_\textrm{\tiny max}}$ (Figure~\ref{20101106BDist} bottom). 

Averaging the magnetic field over the area of an SCB approximates the measurements of SCBs made using MDI. The top row of Figure~\ref{20101106BDist} shows a somewhat similar distribution in each component of the magnetic field to those seen in MDI -- a mostly compact distribution of intensities with average intensities near zero.  The zonal, meridional, and radial directions all had intensities of $2 - 3$ G compared to -5.1 G in MDI. Given the instrumental measurement error in HMI of $\approx 10$ G, these measurements are consistent with each other. 

When examining the vector magnetic fields within an SCB, we observe apparent substructure on the scale of one or two HMI pixels. Selectively examining the magnetic extrema contained within the detected SCB boundary reveals significantly different results than averaging the magnetic intensity. The bottom row of Figure~\ref{20101106BDist} shows the distribution of the magnetic vector components of these extrema.  This analysis of SCB magnetic field components is clearly bifurcated into two distinct distributions in the positive and negative directions for each vector components. Table~\ref{t:HMIresults} lists the number and mean intensity of the magnetic extrema for each of these distributions. 

\begin{table}
  \caption{HMI vector magnetic field measurements over all SCBs for extrema within the measured SCB radius ($\vec{B}_{\phi,B_\textrm{\tiny max}}$, $\vec{B}_{\theta,B_\textrm{\tiny max}}$, and $\vec{B}_{r,B_\textrm{\tiny max}}$). }
\begin{tabular}{llcccccc}     
\dhline
{Event}	& 	&\multicolumn{3}{c}{{Postive Direction}}						& \multicolumn{3}{c}{{Negative Direction}} \\
{Date} 	&	& $\vec{B}_{\phi}$ 	& $\vec{B}_{\theta}$ 	& $\vec{B}_{r}$ 		&  $\vec{B}_{\phi}$ 	& $\vec{B}_{\theta}$ 	& $\vec{B}_{r}$\\
\hline
6 Nov. 2010	&SCB Count				& 134 & 109 & 124		&	76 & 101 & 86  \\
			&Mean ${B}$ (G)		& 57 & 102 & 101		&	 -54 & -101 & -94	\\
			&			&				&				\\
30 Nov. 2010 	&SCB Count				& 20 & 11 & 20		& 	28 & 37 & 28  \\
			&Mean ${B}$ (G)		& 70 & 90 & 59	&	-69 & -99 & -71 \\
\hline
 \end{tabular}  
\label{t:HMIresults}
\end{table}

In both the 6 November and 30 November events, the directional bifurcation of the magnetic vector components in the positive and negative directions are quasi-symmetric (Table~\ref{t:HMIresults}). The number of components in the positive direction outnumber those with negative direction on 6 November, but this trend is reversed in the 30 November event. The largest magnetic intensity in both events is in the meridional direction ($\ge 90$~G), however all directions have mean magnetic intensities above 50 G. This compares to a mean magnetic field magnitude of $<10$~G in all but one event when measured with MDI. 

Combining the results of 6 and 30 November, we can define the vector components of a typical extrema within an SCB with the mean component magnitudes. This typical SCB has a zonal component ($|B_{\phi}|$) magnitude of $58\pm1.4$~G, a meridional component ($|B_{\theta}|$) magnitude of $100\pm1.8$~G, and radial component ($|B_r|$) magnitude of $92\pm1.9$~G. The overall peak magnetic magnitude of this typical SCB ($|B_{SCB}|$) is $148\pm2.9$~G. 

\subsubsection{Modeled Field}
\label{s:bmodeled}

A PFSS model from \citet{Schrijver2003} is used to model the coronal magnetic field above SCBs (see Section~\ref{s:pfss}). We choose to use a PFSS model based upon the magnetic model of SCBs from~\citet{Pevtsov2007}, which describes SCBs originating at the base of already existing field lines. A PFSS model is appropriate to model the magnetic field lines responsible for the location of SCBs because these field lines are in equilibrium given a pre-flare static coronal magnetic field and the footpoints of field lines are determined by the photospheric magnetic field. 

\begin{figure}
   \centerline{\includegraphics[width=0.6\textwidth,clip=0,angle=0]{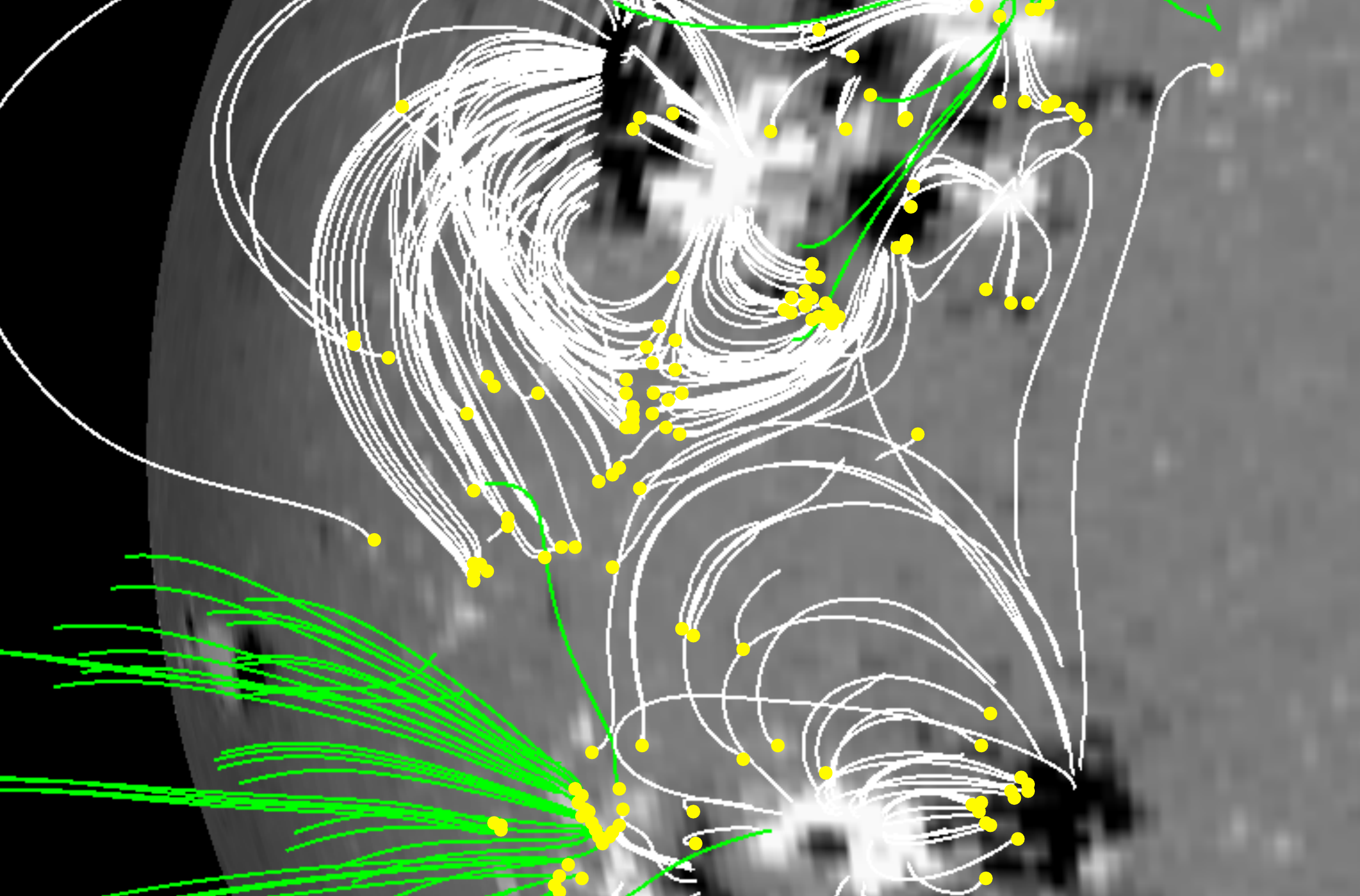}} 

              \caption{An example of the magnetic field lines above SCBs modeled using a PFSS approach from 19 December 2002. The green lines are positive open field lines and closed lines are plotted in white above an MDI magnetogram. One footpoint of each modeled loop is rooted in an SCB. The SCBs visible in this image are indicated with a yellow dot. }
   \label{SCB_PFSS}
   \end{figure}

The locations of SCBs are remapped onto photospheric magnetograms and used as the initialization grid for tracing the potential field lines. This technique produces a localized map of the magnetic field lines around SCBs (Figure~\ref{SCB_PFSS}). The results of this modeling for all events are shown in Table~\ref{t:pfsschar}. An average field line derived from an SCB had a relatively consistent length of $0.20 \pm 0.09\ R_\odot$, and the maximum line length averaged $0.96 \pm 0.32\ R_\odot$ and never exceeded $1.45\ R_\odot$. This model does not describe the dynamics driving the creation of SCBs, but it does give valuable context to the origin and characteristics of their magnetic footpoints.
\begin{table}
  \caption{Characteristics of closed magnetic loops derived by the PFSS model for the photospheric magnetic field beneath SCBs.}
\begin{tabular}{lccc}     
\dhline
 Event &  Mean & Maximum & Minimum \\ 
Date & Length ($R_\odot$) & Length ($R_\odot$) & Length ($R_\odot$)\\
\hline
19 Dec. 2002 & 0.25 & 1.19 & 0.009 \\
6 Mar. 2003 & 0.30 & 1.44 & 0.010\\
9 Mar. 2003 & 0.13 & 0.81 & 0.014\\
11 Jun. 2003 & 0.18 & 0.91 & 0.012\\
29 Oct. 2003 & 0.21 & 1.28 & 0.005\\
9 Nov. 2004 & 0.26 & 0.86 & 0.011\\
6 May 2005 & 0.13 & 0.94 & 0.010\\
13 May 2005 & 0.19 & 1.09 & 0.009\\
6 Dec. 2006 & 0.39 & 1.13 & 0.009\\
6 Nov. 2010 & 0.10 & 0.33 & 0.012\\
30 Nov. 2010 & 0.09 & 0.53 & 0.011\\

 \hline
 \end{tabular}  
\label{t:pfsschar}
\end{table}

Although a PFSS model of the coronal magnetic field is not an accurate predictor for the local magnetic field lines surrounding a dynamically changing eruption for the reasons discussed in Section~\ref{s:pfss}, some general findings can be made from this model. First, over 95\% of the lines modeled were closed, and of the closed lines, none reached beyond $1.45\ R_\odot$.  Second, there is no correlation between field line length and intensity or duration of SCB. This lack of correlation means that the strength of the magnetic field is not driving the intensity or duration of the brightening. Third, only a loose correlation is observed between the distance of the SCB from the flare center and field line length. No long field lines are found close to the flare center, and short field lines are most likely to be found near the flare. 

\section{Discussion and Implications} 
	\label{s:discussion}
	
	Tracking flare kernels through the evolution of an erupting flare is an effective way to characterize the associated evolving active region in spatially resolved images. The sum of these kernels reconstructs the profile of the associated X-ray intensity curve of the flare and individual kernels that persist through a majority of the eruption. The flare kernels dissect the flare into its smallest visibly resolvable components in the \Ha\ images. It is not possible to surmise that a single kernel is tracking a specific flare loop in this study. A thorough investigation and discussion of the dynamics of flares and flare kernels can be found in a forthcoming work. 
	
	SCBs are found to be a specific case of chromospheric compact brightenings that occur in conjunction with flares. The distinct nature of SCBs from other brightenings in \Ha\ arises from their impulsive brightening and decay, rapid propagation and dispersal away from the flare center. Despite the observed propagation, the heated plasma in tracked SCB kernels does not physically progress in any coherent direction. SCBs in aggregate often precede the \Ha\ flare peak and have evolutions that are distinct from the GOES soft X-ray emission, which is a proxy for the thermal component of the flare emission. The temporal coincidence yet separation between flare and SCB evolution suggests that SCBs have an origin that is separate from the \Ha\ flare but is driven by a triggering mechanism that is causally connected to the host flare. 

The most common photospheric magnetic field intensity of SCBs is near 0~G when measured in MDI, yet does not necessarily disprove the prevailing unipolarity hypothesis. The weak relationship between SCBs and measured magnetic intensity is most likely to be the result of the quality of the MDI magnetograms. MDI magnetograms used in this study have half the resolution of the \Ha\ images, meaning that four MDI pixels completely covers the entire area of a typical SCB. Also, since the MDI magnetograms only measure LOS magnetic field, the locations of the SCBs on the solar disk (see Table~\ref{t:events}) have a strong influence on the measured magnetic intensity. The closer the measurements are to the limb, the more tangential the magnetic field measured in the magnetogram and thus closer to zero. A highly inclined magnetic field might also skew measurements, since the magnetic polarity directly beneath the SCB would then not represent the strength of the field line connecting the photosphere to the chromosphere. 

When measuring the vector magnetic components using HMI data, a stronger signal appears (Figure~\ref{SCBMag}). If the HMI magnetic field is averaged over the SCB area, the results are similar to MDI with mean intensities of the SCB population near zero. When only the extrema are selected within the geometric boundaries of the SCB, two distinct populations emerge in the positive and negative directions in each vector component (Figure~\ref{20101106BDist}). The extrema were selected from the strongest measured LOS magnitude of HMI pixels within an SCB boundary without consideration of the vector components of the field. By calculating their skewness and kurtosis, we show the extrema statistical significance.  We demonstrated in Table~\ref{t:HMIresults} the non-triviality that each vector component of those extrema exhibits a bifurcated population. These directional populations are biased in the positive direction on 6 November, but this trend is reversed in the 30 November event. 

The bifurcated populations of magnetic components observed beneath SCBs imply that SCBs are magnetically anchored in the photosphere, {\it i.e.} they are not purely atmospheric phenomena. This finding does not disagree with the prevailing unipolarity hypothesis. Visually distinct magnetic features are also routinely observed within the boundary of a given SCB, which suggests that the magnetic footpoint of an SCB in the photosphere is significantly smaller than the area of the \Ha\ chromospheric brightening. Higher resolution \Ha\ imaging of SCBs is needed to more fully investigate the apparent size discrepancy between the photospheric and chromospheric measurements. 

There is a lack of any significant correlation between the distance at which SCBs appear and modeled magnetic field line length, suggesting that the heuristic model presented by \citet{Kirk2012a} is most likely an oversimplified schematic of the eruption. This heuristic model postulates that SCBs arise from a series of nested quiescent coronal loops that are disturbed when the flare erupts. However, a majority of magnetically modeled SCBs have field lines that exist between $0.1$ and $0.3\ R_\odot$ above the photosphere, indicating that the mechanism for driving SCBs is most likely in the lower corona.

 \begin{figure}
   \centerline{\includegraphics[width=0.95\textwidth,clip=0,angle=0]{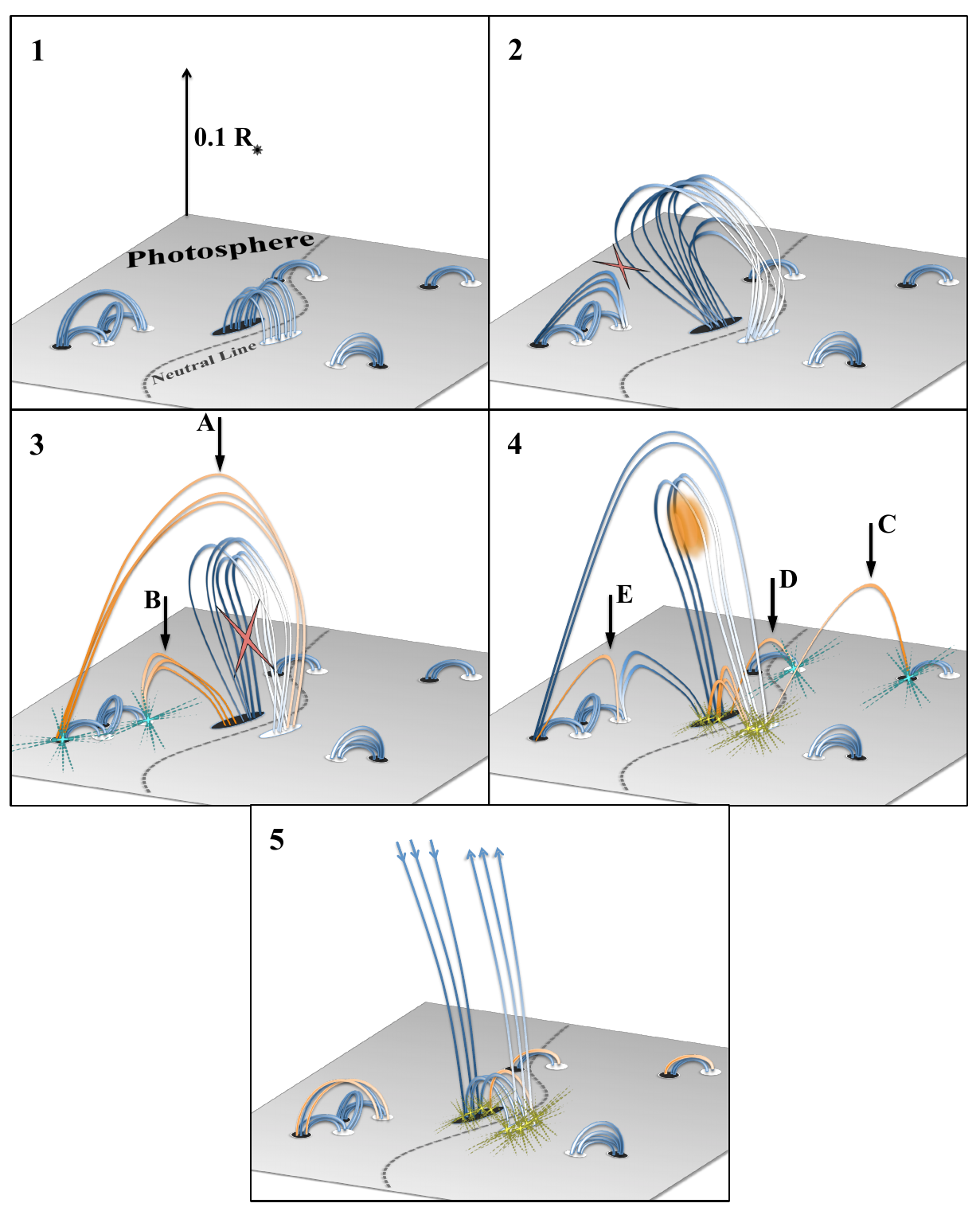}} 

              \caption{A heuristic interpretation of the temporal progression of SCBs and the associated flare. Magnetic loops, shown in blue, connect negative and positive polarity photospheric footpoints in the photosphere, shown in black and white. Newly reconnected magnetic field lines are highlighted in orange. A red ``X'' indicates sites of magnetic reconnection and diffraction spikes in green or yellow indicate regions of observed intensity enhancement of SCBs or flare ribbons respectively.  }
   \label{SCB_diagram}
   \end{figure}

Figure~\ref{SCB_diagram} illustrates our current understanding of the temporal progression of SCBs and eruption of the associated flare. Panel 1 depicts a potential-field solution to the photospheric magnetogram of the flaring region, such as performed by the PFSS model. The flaring region is shown in the center as well as a quadrupolar active region and three small dipolar field configurations. 

Panel 2 in Figure~\ref{SCB_diagram} shows the initial destabilization of the flaring region and the accompanying magnetic field lines expanding into the localized area and the reaction of the quadrupolar region. The red ``X'' indicates the location where reconnection takes place between the flare magnetic field and the quadrupolar active region. 

Panel 3 in Figure~\ref{SCB_diagram} shows the flaring region before flare reconnection but after the appearance of the first SCBs. Loops marked {\bf A} and {\bf B} are newly reconnected from the x-point in panel 2 and the SCBs observed as a result of plasma accelerated by this reconnection are shown by the green diffraction spikes. The loops marked as {\bf A} are now overarching the entire erupting region and confine the expanding eruption. The red ``X'' denotes where the main flare reconnection takes place. 

Panel 4 in Figure~\ref{SCB_diagram} shows the flare arcade loops as well as the erupting flux rope, both in orange. The observed flare ribbons are indicated with yellow diffraction spikes. As the overarching magnetic field lines ({\bf A}) are further stressed, they become unstable themselves and reconnect (not shown) to magnetic footpoints further away from the flare. These newly connected lines, marked as lines {\bf C} and {\bf D}, also produce observable SCBs indicated by green diffraction spikes. As the flare eruption evolves it also releases some magnetic stress imposed on the quatrupolar region, allowing a more potential field to reemerge as indicated by line {\bf E}. 

Panel 5 in Figure~\ref{SCB_diagram} shows the flux rope eruption moving beyond the local region depicted and the dipolar and quadrupolar regions relaxing into a potential field configuration. Notice that one dipolar region remained unchanged throughout the eruption. This is due to the physical geometry of the eruption itself. Two ribbon flares have an elongated axis that makes it favorable for field lines to connect to regions tangentially aligned with this axis. Regions perpendicular to this flare axis are more resilient to reconnection with the flare arcade and thus less prone to SCBs. 
		
	The findings presented so far all constrain the triggering mechanism for SCBs. Given the likely presence of a CME when SCBs are observed, we can postulate that both come from a common physical origin in the lower corona. Directly driven models of CME release are inconsistent with the measurements of SCBs made here. In a thermal blast model~\citep[\emph{e.g.}][]{Dryer1982}, thermal pressure is the driver of a CME. In that case, we would expect to see a correlation between the strength and propagation velocity of SCBs to scale with the energy release of flares. There is no such connection between the energy released in the flare and propagation speed of SCBs. 
	
In the CME breakout model, \citet{Antiochos1999} postulate that the energy stored in a closed and sheered arcade plays a crucial role in determining when a CME eruption will occur. In this model, the reconnection event that triggers a CME is a low-energy event located above the erupting arcade. This low-energy pre-CME reconnection event at first glance seems like it could be the triggering mechanism for SCBs since SCBs are the result of accelerated particles impacting the chromosphere~\citep{Kirk2014} and often occur before the onset of the primary flare emission. If this pre-CME reconnection described SCBs, we would expect to observe a correlation between the measured physical parameters of SCBs (\emph{e.g.} number of SCBs and \Ha\ intensity) and the physical characteristics of the associated CME (\emph{e.g.} velocity and mass). For example, in this scenario if a high speed CME were released, there must be either a high energy pre-CME reconnection or many smaller reconnections and we would expect to observe either more SCBs or SCBs with higher \Ha\ intensities. We do not observe any correlation between the plasma dynamics of SCBs and the measured properties of the associated CME. This lack of correlation implies that SCBs are not a direct product of the driver of CME release.


\section{Conclusions} 
	\label{s:conclusions}
	Sequential coronal brightenings (SCBs) are a type of localized chromospheric heating and ablation due to impacting coronal plasma~\citep{Kirk_PhD,Kirk2014}. SCBs are unlike flare ribbons in that they are a secondary effect of solar eruptions. Using an automated detection and tracking algorithm, we can now readdress the three questions asked at the outset of this investigation. (1) Is there a relationship between the properties of the host flare and associated SCBs? (2) Is there a consistent photospheric magnetic field component to SCBs? (3) Can a potential coronal field model help explain the other SCB measurements? 
	
Addressing the first question, SCBs originate in quiescent coronal magnetic loops above a chromospheric flare that are forcibly disturbed in the course of the flare eruption. As these coronal tethers reconfigure, trapped plasma is now free to cascade into the chromosphere causing SCBs. This model is affirmed by the finding that 82\% (9 out of 11) of SCB events in this study accompany subsequent visible CMEs, also validating the conclusions of~\citet{Bala2006}. The corona and chromosphere are visibly disturbed by each other in the eruption process of a filament or flare and the progressive trains of SCBs observed propagating away from the flare with velocities in two distinct groups: a slow group with speeds between 35 and 85~\kms, and a fast group propagating at speeds $>150$~\kms. These groups of SCBs are evidence of the pre-flare surrounding magnetic topology rather than the flare itself. 

SCBs are more closely related to the release of an associated CME rather than the flare itself. Viewing SCBs as a product of CME formation and eruption explains why flares releasing vastly different energies (from no X-ray signature to an X10.0 flare) have SCBs with similar physical traits. Since SCBs can first appear before the onset of \Ha\ emission in a flare, during the impulsive phase, or after the peak intensity, an associated CME is most likely to be the process driving the appearance of SCBs rather than flare reconnection. Chromospheric observations of flare and filament eruptions (such as with ISOON) can give an indication of an emerging CME up to a couple hours before the CME is detected with a coronagraph. 

The second question asked by this study relates to the magnetic flux of SCBs. Several other works on SCBs state that they should be unipolar~\citep[\emph{e.g.}][]{Bala2006,Kirk2012a}. This work found a slight magnetic bias when measured with MDI. When measuring magnetic vector magnetic field components of SCBs using HMI, a significant signal is observed. Each vector magnetic component in both events with HMI coverage showed a distinctly nonzero result in either the positive or negative direction. These findings suggest a magnetic footpoint of SCBs in the photosphere with a diameter significantly smaller than that in the chromospheric brightening with a peak intensity of $148\pm2.9$~G.

Lastly, we explored a potential field model of SCBs. A potential model seems illogical at first, since SCBs are dynamic features which are most certainly non-potential. But because SCBs are more indicative of the quiescent pre-flare environment than the flare itself, a potential model makes sense as a first-order estimate as to where SCBs form. A PFSS model suggests that the origin of SCBs lies in the lower corona, at heights around $0.1\ R_\odot$ above the solar surface.

This research finds statistically significant magnetic substructure within the \Ha\ boundary of SCBs. The potential field extrapolation of the magnetic field within SCBs shows magnetic field lines extending into the low corona. Both of these findings combined, point to the conclusion that a single SCB represents a magnetic footpoint of a coronal magnetic loop, which supports the hypothesis that SCBs are unipolar.  It is conceivable that future observations of SCBs will reveal a more complicated substructure or more subtle evolution.

 Lasting only a few minutes, SCBs are fleeting indicators of the solar flare environment at the time of flare eruption. Unfortunately due to their small physical size, low intensity, and short lifetime, the physical description of SCBs is incomplete and hyperspectral imagery of this phenomenon is required for a complete picture.  In all previous studies, SCBs are identified by their Doppler intensity profile in \Ha\ images. The ISOON telescope was the only full disk solar imager that regularly recorded full-disk off-band images in the \Ha\ line with suitable resolution to capture SCBs. Unfortunately this telescope was decommissioned in 2011 and has not been replaced. 

%

%

%
 \begin{acks}
The authors would like to recognize: (i) USAF/AFRL Space Scholar Program, (ii) NSO/AURA for the use of their Sunspot, NM facilities, (iii) AFRL/RVBXS, (iv) NMSU, and (v) the manuscript referee for his/her insightful comments about this work.  MSK was supported by an appointment to the NASA Postdoctoral Program at the Goddard Space Flight Center, administered by Universities Space Research Association through a contract with NASA. KSB was supported by an AFOSR Task: ``Physics of Coupled Flares and CME Systems.''\\
 \end{acks}

{\small \noindent
{\bf Disclosure of Potential Conflicts of Interest}   The authors declare that they have no conflicts of interest.}

%
\bibliographystyle{spr-mp-sola}

\end{article} 
\end{document}